\documentclass[11pt]{article}

\usepackage{amsfonts,amsthm,amssymb}
\usepackage{mathtools}
\usepackage{subcaption}
\usepackage{caption}
\usepackage{url}
\usepackage{mathtools}
\usepackage{todonotes}

\newcommand{\ket}[1]{| #1 \rangle}
\newcommand{\bra}[1]{\langle #1 |}
\newcommand{\braket}[2]{\langle #1 | #2 \rangle}

\newcommand{\neswarrow}{\mathrel{\text{\ooalign{$\swarrow$\cr$\nearrow$}}}}
\newcommand{\nwsearrow}{\mathrel{\text{\ooalign{$\nwarrow$\cr$\searrow$}}}}

\def\UA{\uparrow}
\def\DA{\downarrow}
\def\LA{\leftarrow}
\def\RA{\rightarrow}
\def\SA{\circlearrowleft}
\def\ULA{\nwarrow}
\def\DLA{\swarrow}
\def\URA{\nearrow}
\def\DRA{\searrow}
\def\LRA{\leftrightarrow}
\def\DRULA{\neswarrow}
\def\DLURA{\nwsearrow}

\newcommand{\comment}[1]{}

\title{Lackadaisical quantum walks on triangular and honeycomb 2D grids}

\author{Nikolajs Nahimovs}
\date{\small{Center for Quantum Computer Science, Faculty of Computing, University of Latvia} \\ 
\small{Raina bulv. 19, Riga, LV-1586, Latvia}\\
\small{\texttt{nikolajs.nahimovs@lu.lv}}}

\begin{document}

\maketitle


\begin{abstract}

\noindent
In the typical model, a discrete-time coined quantum walk search has the same running time of $O(\sqrt{N} \log{N})$ for 2D rectangular~\cite{Ambainis:2005}, triangular~\cite{Abal:2012} and honeycomb~\cite{Abal:2010} grids.
It is known that for 2D rectangular grid the running time can be improved to $O(\sqrt{N \log{N}})$ using several different techniques.
One of such techniques is adding a self-loop of weight $4/N$ to each vertex (i.e. making the walk lackadaisical)~\cite{Wong:2018,Hoyer:2020}.

In this paper we apply lackadaisical approach to quantum walk search on triangular and honeycomb 2D grids. We show that for both types of grids adding a self-loop of weight $6/N$ and $3/N$ for triangular and honeycomb grids, respectively, results in $O(\sqrt{N \log{N}})$ running time.

\end{abstract}

 
\section{Introduction}

Quantum walks are quantum counterparts of classical random walks \cite{Portugal:2013}. 
Similarly to classical random walks, there are two types of quantum walks: discrete-time quantum walks (DTQW),  introduced by Aharonov~{\it et al.}~\cite{Aharonov:1993}, and continuous-time quantum walks (CTQW), introduced by Farhi~{\it et al.}~\cite{Farhi:1998}.
For the discrete-time version, the step of the quantum walk is usually given by two operators -- coin and shift -- which are applied repeatedly. 
The coin operator acts on the internal state of the walker and rearranges the amplitudes of going to adjacent vertices. The shift operator moves the walker between the adjacent vertices.

Quantum walks have been useful for designing algorithms for a variety of search problems\cite{Nagaj:2011}.
To solve a search problem using quantum walks, we introduce the notion of marked elements (vertices), corresponding to elements of the search space that we want to find.
We perform a quantum walk on the search space with one transition rule at the unmarked vertices, and another transition rule at the marked vertices. If this process is set up properly, it leads to a quantum state in which the marked vertices have higher probability than the unmarked ones. This method of search using quantum walks was first introduced in \cite{Shenvi:2003} and has been used many times since then.

The problem of search on a two-dimensional rectangular grid was stated in 2002 by Paul Benioff \cite{Benioff:2002}, who conjectured that local search on $\sqrt{N} \times \sqrt{N}$ grid needs $\Omega(N)$ time, i.e. no quantum speed-up is possible.
One year later Ambainis and Aaronson proposed an algorithm \cite{Aaronson:2003} which finds a marked vertex in $O(\sqrt{N} \log^2{N})$ steps.
In 2005 Ambainis, Kempe and Rivosh \cite{Ambainis:2005} proposed a quantum walk based algorithm (AKR algorithm) which finds a marked vertex with $O(\frac{1}{\log{N}})$ probability in $O(\sqrt{N\log N})$ steps. Applying amplitude amplification this gives the running time of $O(\sqrt{N}\log N)$. 
Following the AKR algorithm, it had been conjectured that the running time can be reduced to $O(\sqrt{N\log N})$, hence, providing a full quadratic speed-up over the random walk based approach.
This conjecture has been confirmed a few years later by Tulsi who in 2008 showed how to modify the AKR algorithm to achieve a constant success probability in $O(\sqrt{N\log N})$ steps \cite{Tulsi:2008}. 

Another method to achieve $O(\sqrt{N\log N})$ running time is to make the quantum walk lackadaisical, i.e. to add a self-loop to each vertex\footnote{There are also other methods, e.g. to run ARK algorithm and to classically search the neighbourhood of a found vertex~\cite{Nahimovs:2013}}.
The concept of lackadaisical quantum walk (quantum walk with self loops) was first studied for DTQW on one-dimensional line \cite{Norio:2005,Stefanak:2014} and later applied to improve the DTQW based search on the complete graph~\cite{Wong:2015a} and two-dimensional rectangular grid~\cite{Wong:2018,Hoyer:2020}.
The running time of the lackadaisical walk heavily depends on a weight of the self-loop. For a rectangular 2D grid with a single marked vertex one optimal weight of the self-loop is $4/N$.

Following the AKR algorithm for the rectangular 2D grid, Abal {\it et.al.} studied the AKR walk on honeycomb~\cite{Abal:2010} and triangular~\cite{Abal:2012} grids. They showed that the walk has the same running time of $O(\sqrt{N} \log N)$, which by applying the Tulsi modification can be reduced to $O(\sqrt{N\log N})$.

In this paper we apply lackadaisical approach to quantum walk search on triangular and honeycomb 2D grids. We show that for both types of grids adding a self-loop of weight $6/N$ and $3/N$ for triangular and honeycomb grids, respectively, results in $O(\sqrt{N \log{N}})$ running time, i.e. in the same improvement which is achieved by using the Tulsi modification.



\section{Quantum walks on the two-dimensional grid}\label{sec:definitions}

\subsection{Non-lackadaisical quantum walk on triangular 2D grid}

Consider a two-dimensional triangular grid of $N$ vertices with periodic boundary conditions.
This defines the $N$-dimensional Hilbert space ${\cal{H_P}}$ associated with the position. 
The coin subspace of the walk is 6-dimensional Hilbert space spanned by the set of states $\{\ket{c}: c\in \{\ULA,\URA,\LA,\RA,\DLA,\DRA\}\}$. The  Hilbert space of the quantum walk is $\mathbb{C}^N\otimes \mathbb{C}^6$.

\begin{figure}[!htb]
\centering
\subcaptionbox{Two-dimensional triangular grid.}{\includegraphics[scale=0.45]{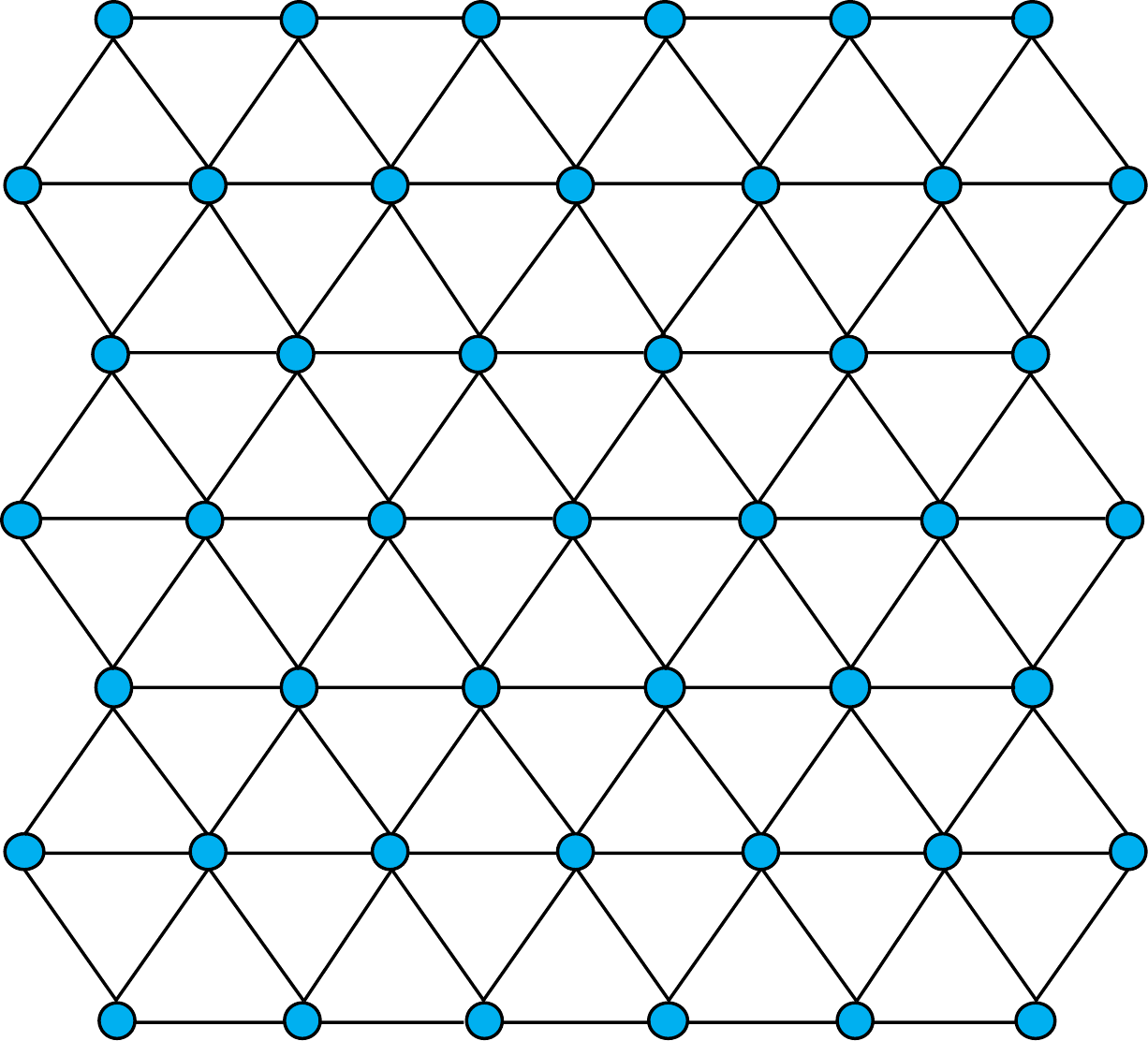}}
\hfill
\subcaptionbox{Triangular grid mapped to rectangular grid.}{\includegraphics[scale=0.45]{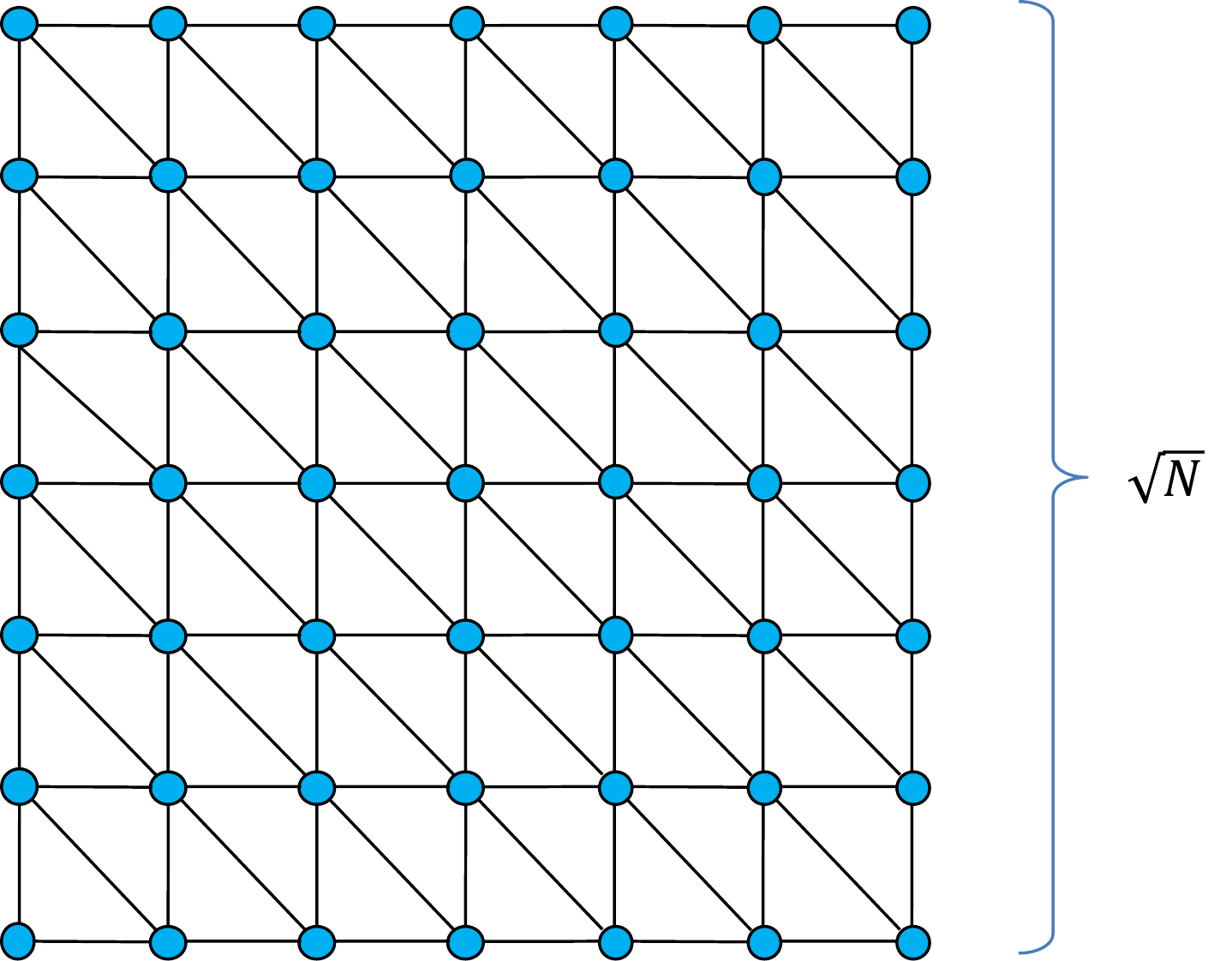}}
\label{fig:tri_on_rect}
\caption{Two-dimensional triangular grid and its mapping to rectangular grid}
\end{figure}

The evolution of a state of the walk (without searching) is driven by the unitary operator $U = S\cdot (I_N \otimes C)$, where $S$ is the flip-flop shift operator and $C$ is the coin operator, given by the Grover's diffusion transformation 
\begin{equation}
C = 2 \ket{s_c}\bra{s_c} - I_6
\end{equation}
with 
$$
\ket{s_c} = \frac{1}{\sqrt{6}}(\ket{\ULA} + \ket{\URA} + \ket{\LA} + \ket{\RA} + \ket{\DLA} + \ket{\DRA}) .
$$

\begin{figure}[!htb]
\centering
\includegraphics[scale=0.6]{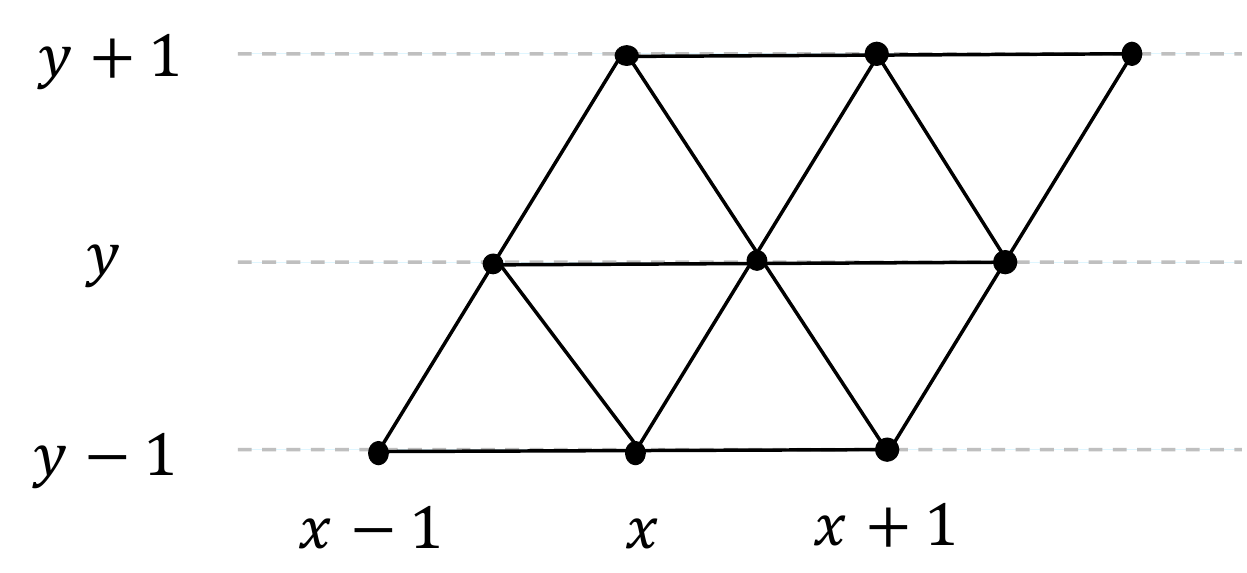}
\caption{Mapping from triangular grid to rectangular grid.}
\label{fig:tri_to_rect_map}
\end{figure}

There exists a simple mapping from triangular to rectangular grid as shown on figure \ref{fig:tri_to_rect_map}. 
This allows us to label the locations of the grid by the coordinates $(x,y)$ for $x, y \in \{0,\dots,\sqrt{N}-1\}$.
The flip-flop shift operator $S$ then can be written as
\begin{eqnarray}
S\ket{x,y,\ULA} & = & \ket{x-1,y+1,\DRA} \\
S\ket{x,y,\DRA} & = & \ket{x+1,y-1,\ULA} \nonumber \\
S\ket{x,y,\LA} & = & \ket{x-1,y,\RA} \nonumber \\
S\ket{x,y,\RA} & = & \ket{x+1,y,\LA}, \nonumber \\
S\ket{x,y,\DLA} & = & \ket{x,y-1,\URA} \nonumber \\
S\ket{x,y,\URA} & = & \ket{x,y+1,\DLA} \nonumber 
\end{eqnarray}

The system starts in 
\begin{equation}
\ket{\psi(0)} = \frac{1}{\sqrt{N}} \sum_{x,y=0}^{\sqrt{N}-1} \ket{x,y} \otimes \ket{s_c} ,
\end{equation}
which is uniform distribution over vertices and directions. Note, that this is a unique eigenvector of $U$ with eigenvalue $1$. The state of the system after $t$ steps is $\ket{\psi(t)} = U^t \ket{\psi(0)}$.

To use quantum walk as a tool for search, we extend the step of the algorithm, making it
$$
U' = U \cdot (Q \otimes I_6) ,
$$
where $Q$ is the query transformation which flips the sign at a marked vertex, irrespective of the coin state. 
Note that $\ket{\psi(0)}$ is a 1-eigenvector of $U$ but not of $U'$.
If there are marked vertices, the state of the algorithm starts to deviate from $\ket{\psi(0)}$.
In case of a single marked vertex, similar to rectangular grid case, after $O(\sqrt{N\log{N}})$ steps the inner product $\braket{\psi(t)}{\psi(0)}$ becomes close to $0$.
If the state is measured at this moment, the probability of finding a marked vertex is $O(1 / \log{N})$~\cite{Abal:2012}.
With amplitude amplification this gives the total running time of $O(\sqrt{N} \log{N})$ steps.


\subsection{Lackadaisical quantum walk on triangular 2D grid}

In case of lackadaisical quantum walk the coin subspace of the walk is 7-dimensional Hilbert space spanned by the set of states $\{\ket{c}: c\in \{\ULA, \URA, \LA, \RA, \DLA, \DRA, \SA \}\}$. The  Hilbert space of the quantum walk is $\mathbb{C}^N\otimes \mathbb{C}^7$.

The shift operator acts on a self loop as 
\begin{equation}
S\ket{x,y,\SA} = \ket{x,y,\SA} .
\end{equation}
The coin operator is 
\begin{equation}
C = 2 \ket{s_c}\bra{s_c} - I_7
\end{equation}
with 
$$
\ket{s_c} = \frac{1}{\sqrt{6 + l}}(\ket{\ULA} + \ket{\URA} + \ket{\LA} + \ket{\RA} + \ket{\DLA} + \ket{\DRA} + \sqrt{l}\ket{\SA}) .
$$
The system starts in 
\begin{equation}
\ket{\psi(0)} = \frac{1}{\sqrt{N}} \sum_{x,y=0}^{\sqrt{N}-1} \ket{x,y} \otimes \ket{s_c} ,
\end{equation}
which is uniform distribution over vertices, but not directions. As before $\ket{\psi(0)}$ is a unique 1-eigenvector of $U$.

In case of search the step of the algorithm is $U' = U \cdot (Q \otimes I_7)$.
In the next sections we will investigate the optimal weight of the self-loop and the running time of the search algorithm.


\subsection{Non-lackadaisical quantum walk on honeycomb 2D grid}

Consider a two-dimensional triangular grid of $N$ vertices with periodic boundary conditions.
This defines the $N$-dimensional Hilbert space ${\cal{H_P}}$ associated with the position. 
The coin subspace of the walk is 6-dimensional Hilbert space spanned by the set of states $\{\ket{c}: c\in \{\ULA,\URA,\LA,\RA,\DLA,\DRA\}\}$. There are two types of vertices, having either $\{\ULA,\DLA,\RA\}$ or $\{\LA,\URA,\DRA\}$ directions. Therefore, instead of $6$ dimensional coin space we can use $3$ dimensional coin space $\{\LRA,\DRULA,\DLURA\}$ which corresponds to $\{\RA,\ULA,\DLA\}$ or $\{\LA,\DRA,\URA\}$ depending on the type of a vertex.
The Hilbert space of the quantum walk is $\mathbb{C}^N\otimes \mathbb{C}^3$.

\begin{figure}[!htb]
\centering
\subcaptionbox{Two-dimensional hexagonal grid.}{\includegraphics[scale=0.35]{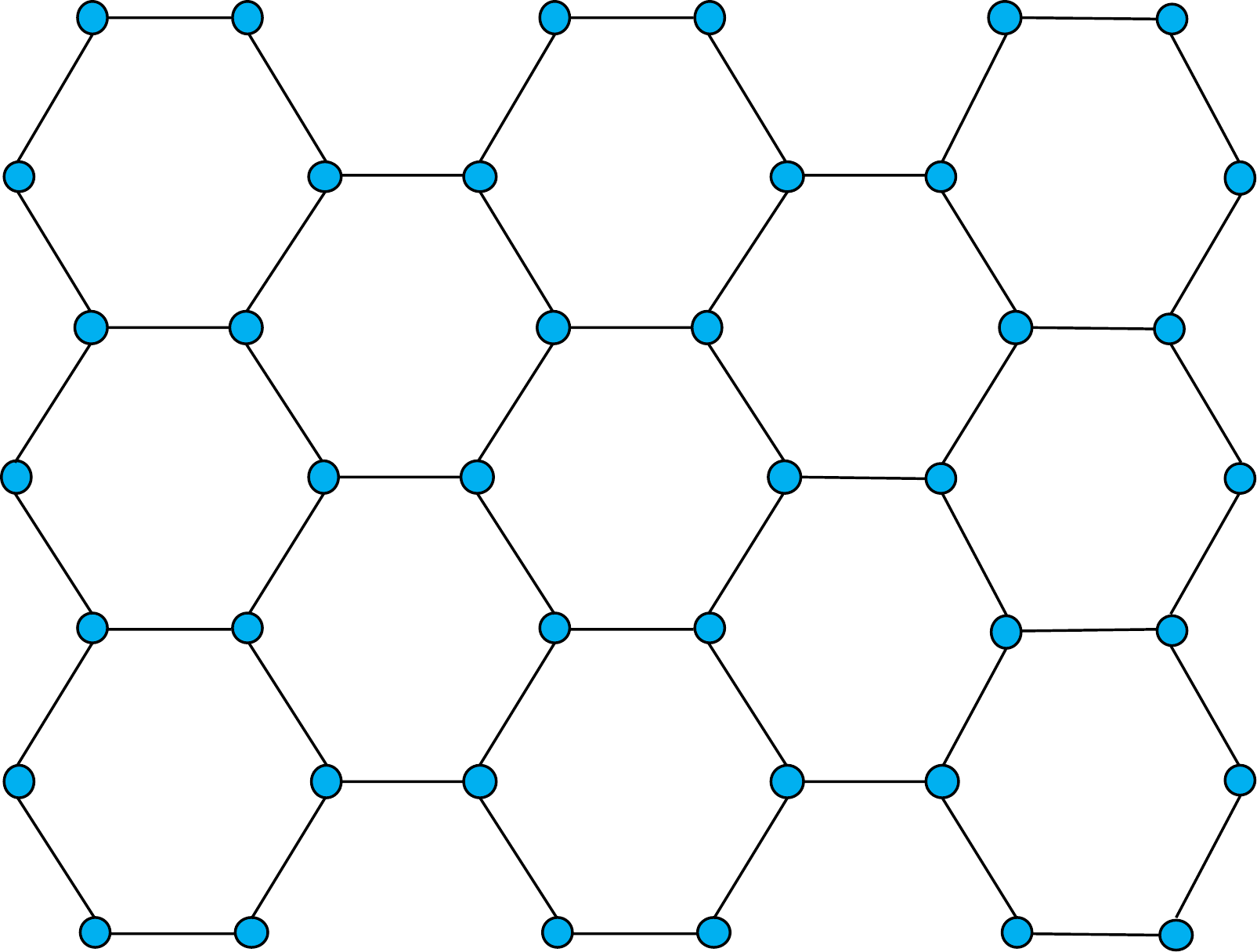}}
\hfill
\subcaptionbox{Hexagonal grid mapped to rectangular grid.}{\includegraphics[scale=0.4]{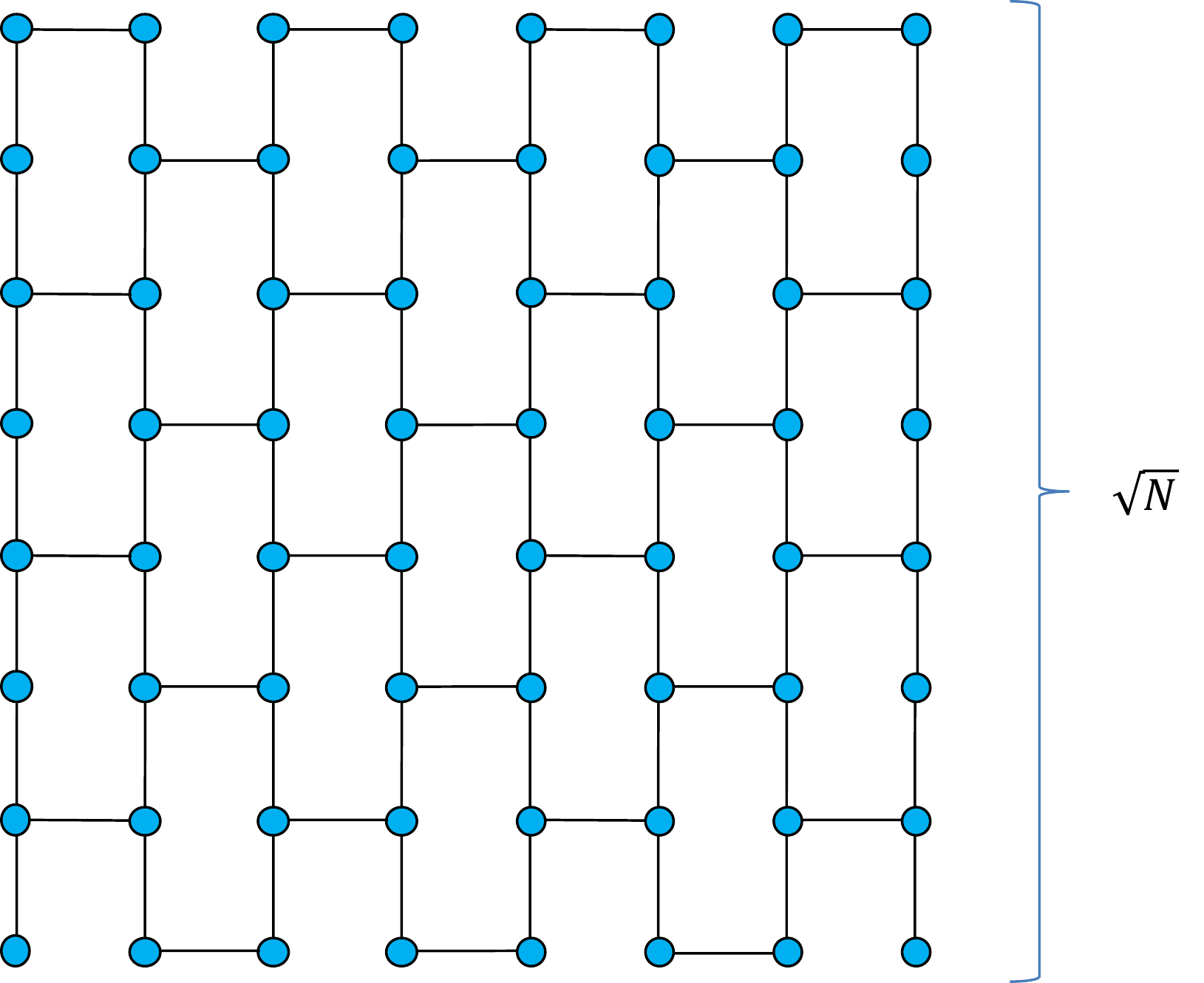}}
\label{fig:hex_on_rect}
\end{figure}

The evolution of a state of the walk (without searching) is driven by the unitary operator $U = S\cdot (I_N \otimes C)$, where $S$ is the flip-flop shift operator and $C$ is the coin operator, given by the Grover's diffusion transformation
\begin{equation}
C = 2 \ket{s_c}\bra{s_c} - I_3
\end{equation}
with 
$$
\ket{s_c} = \frac{1}{\sqrt{3}}(\ket{\LRA} + \ket{\DLURA} + \ket{\DRULA}).
$$

\begin{figure}[!htb]
\centering
\includegraphics[scale=0.6]{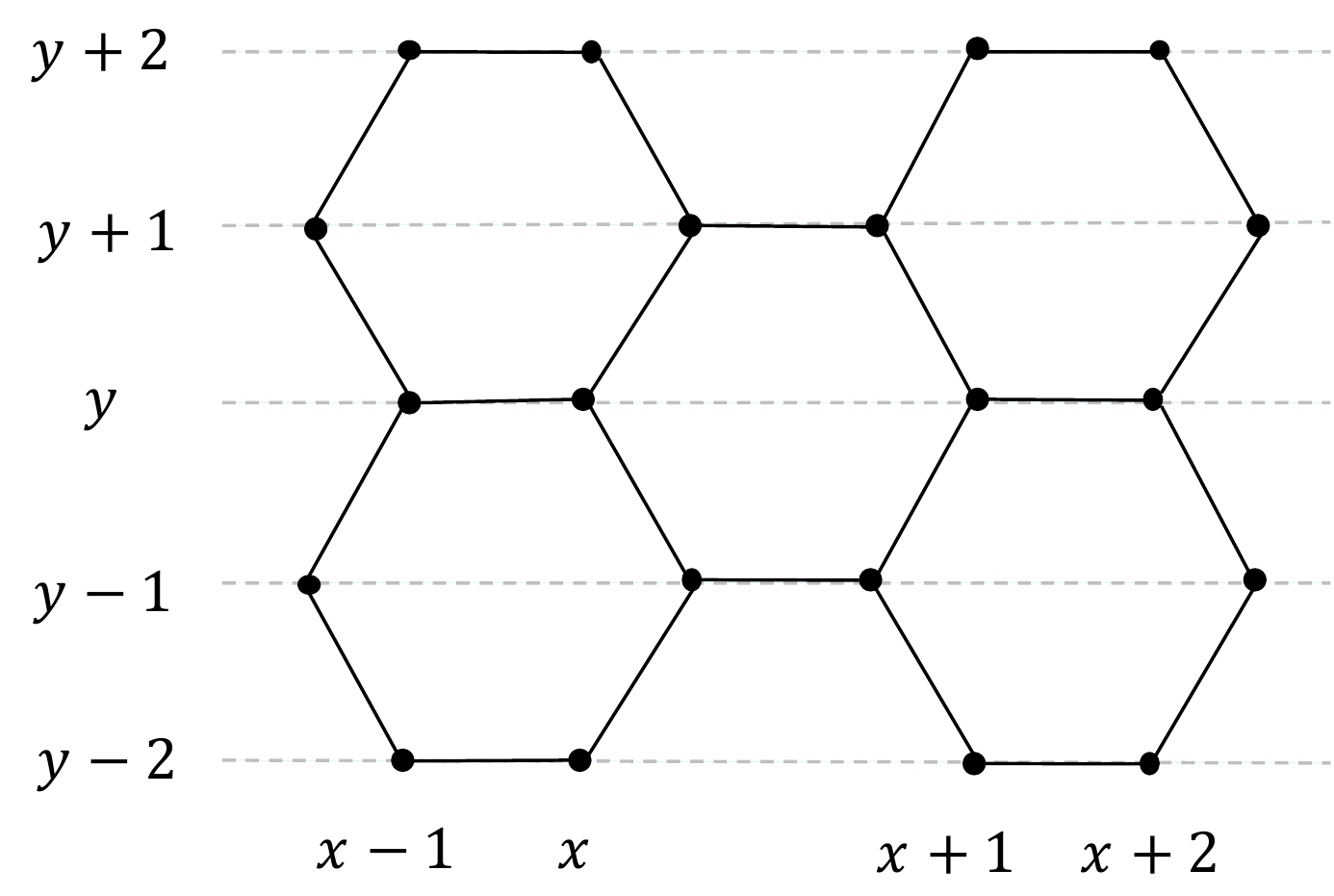}
\caption{Mapping from hexagonal grid to rectangular grid.}
\label{fig:hex_to_rect_map}
\end{figure}

There exists a simple mapping from hexagonal to rectangular grid as shown on figure \ref{fig:hex_to_rect_map}. 

\noindent
This allows us to label the locations of the grid by the coordinates $(x,y)$ for $x, y \in \{0,\dots,\sqrt{N}-1\}$.
The flip-flop shift operator $S$ then can be written as
\begin{eqnarray}
S\ket{x,y,\ULA} & = & \ket{x,y+1,\DRA} \\
S\ket{x,y,\DRA} & = & \ket{x,y-1,\ULA} \nonumber \\
S\ket{x,y,\LA} & = & \ket{x-1,y,\RA} \nonumber \\
S\ket{x,y,\RA} & = & \ket{x+1,y,\LA}, \nonumber \\
S\ket{x,y,\DLA} & = & \ket{x,y-1,\URA} \nonumber \\
S\ket{x,y,\URA} & = & \ket{x,y+1,\DLA} \nonumber 
\end{eqnarray}

The system starts in 
\begin{equation}
\ket{\psi(0)} = \frac{1}{\sqrt{N}} \sum_{x,y=0}^{\sqrt{N}-1} \ket{x,y} \otimes \ket{s_c} ,
\end{equation}
which is uniform distribution over vertices and directions. Note, that this is a unique eigenvector of $U$ with eigenvalue $1$. The state of the system after $t$ steps is $\ket{\psi(t)} = U^t \ket{\psi(0)}$.

To use quantum walk as a tool for search, we extend the step of the algorithm, making it
$$
U' = U \cdot (Q \otimes I_3) ,
$$
where $Q$ is the query transformation which flips the sign at a marked vertex, irrespective of the coin state. 
Note that $\ket{\psi(0)}$ is a 1-eigenvector of $U$ but not of $U'$.
If there are marked vertices, the state of the algorithm starts to deviate from $\ket{\psi(0)}$.
In case of a single marked vertex, similar to rectangular grid case, after $O(\sqrt{N\log{N}})$ steps the inner product $\braket{\psi(t)}{\psi(0)}$ becomes close to $0$.
If the state is measured at this moment, the probability of finding a marked vertex is $O(1 / \log{N})$~\cite{Abal:2010}.
With amplitude amplification this gives the total running time of $O(\sqrt{N} \log{N})$ steps.


\subsection{Lackadaisical quantum walk on honeycomb 2D grid}

In case of lackadaisical quantum walk the coin subspace of the walk is 4-dimensional Hilbert space spanned by the set of states $\{\ket{c}: c\in \{\LRA, \DRULA, \DLURA, \SA \}\}$. The  Hilbert space of the quantum walk is $\mathbb{C}^N\otimes \mathbb{C}^4$.

The shift operator acts on a self loop as 
\begin{equation}
S\ket{x,y,\SA} = \ket{x,y,\SA} .
\end{equation}
The coin operator is 
\begin{equation}
C = 2 \ket{s_c}\bra{s_c} - I_4
\end{equation}
with 
$$
\ket{s_c} = \frac{1}{\sqrt{3 + l}}(\ket{\LRA} + \ket{\DLURA} + \ket{\DRULA} + \sqrt{l}\ket{\SA}) .
$$
The system starts in 
\begin{equation}
\ket{\psi(0)} = \frac{1}{\sqrt{N}} \sum_{x,y=0}^{\sqrt{N}-1} \ket{x,y} \otimes \ket{s_c} ,
\end{equation}
which is uniform distribution over vertices, but not directions. As before $\ket{\psi(0)}$ is a unique 1-eigenvector of $U$.

In case of search the step of the algorithm is $U' = U \cdot (Q \otimes I_4)$.
In the next sections we will investigate the optimal weight of the self-loop and the running time of the search algorithm.


\section{Analysis}\label{sec:analysis}

The running time of the search by the lackadaisical quantum walk heavily depends on a weight of the self-loop. 
It was shown~\cite{Wong:2018} that for two-dimensional rectangular grid of size $\sqrt{N} \times \sqrt{N}$ with a single marked vertex the optimal weight of the self-loop is $l = 4/N$. 
In this section we study search by the lackadaisical quantum walk for a single marked vertex on two-dimensional triangular and honeycomb grids. 
We find optimal weights of the self-loop as well as the total running time of the search algorithm.
The presented data is obtained from numerical simulations.
The .NET code used to simulate the quantum walk search algorithms is available in \cite{Simulator}.



\subsection{Lackadaisical quantum walk on triangular 2D grid}

\begin{figure}[!htb]
\centering
\includegraphics[scale=0.5]{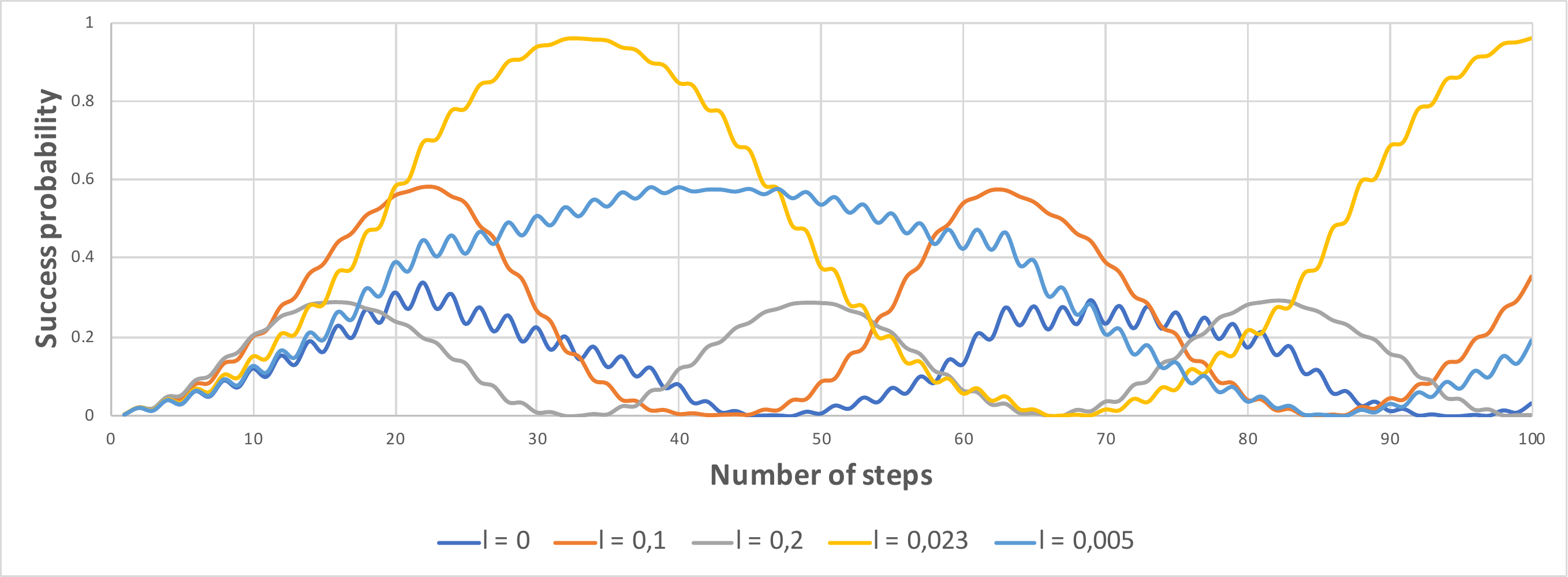}
\caption{Success probability as a function of time for different $l$ on triangular 2D grid of size $16 \times 16$.}
\label{fig:tri_diff_l}
\end{figure}

The running time of the lackadaisical quantum walk depends on a weight of the self-loop $l$. 
Figure \ref{fig:tri_diff_l} shows the evolution of the probability of finding a marked vertex for the lackadaisical quantum walk on a triangular grid of size $N = 16 \times 16$ for various values of $l$. As one can see different values of $l$ result in different success probabilities and numbers of steps till the first peak.

\begin{figure}[!htb]
\centering
\includegraphics[scale=0.5]{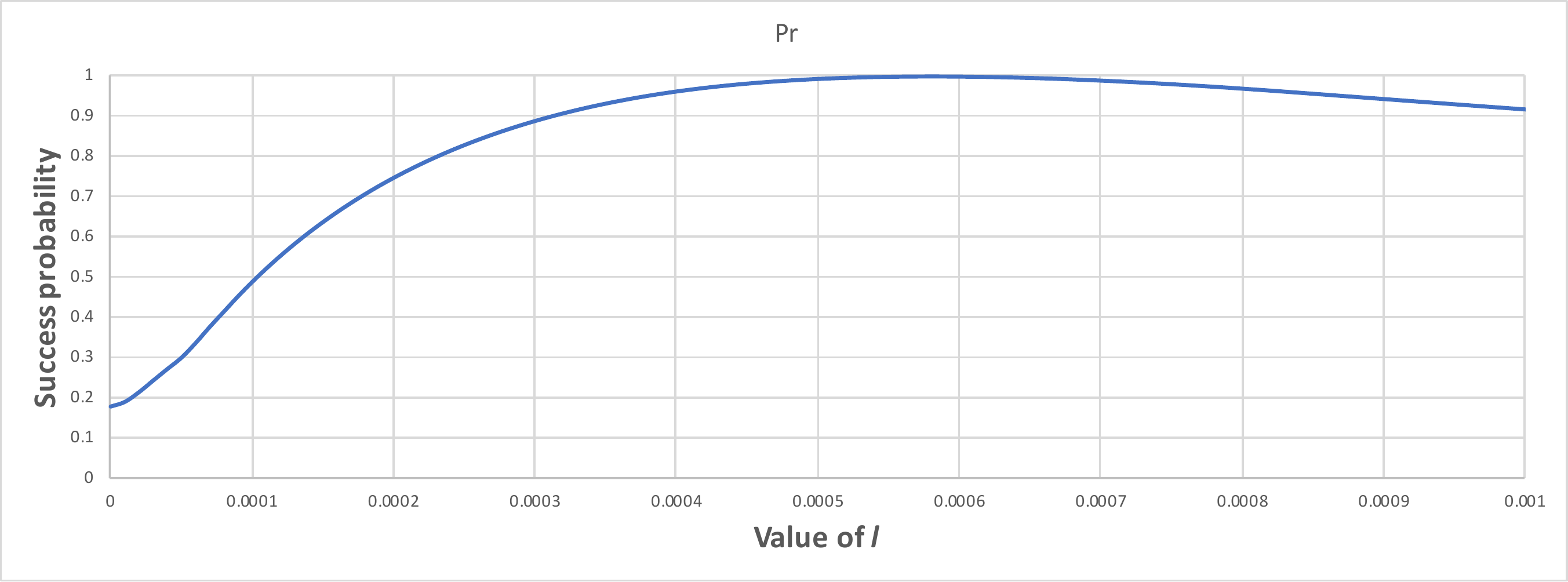}
\caption{Success probability as a function of time for different weights of the self-loop for triangular 2D grid of size $16 \times 16$.}
\label{fig:tri_opt_l}
\end{figure}

Figure \ref{fig:tri_opt_l} shows the success probability for different values of $l$ for search on a triangular grid of size $100 \times 100$.

\noindent
When $l = 0$, the lackadaisical walk reproduces the regular (non-lackadaisical) quantum walk, where the success probability reaches a value of $O(1/\log{N})$ at $O(\sqrt{N\log{N}})$. As $l$ increases, the success probability grows, almost reaching $1$ when $l \approx 0.0006$ which is approximately $6/N$. As $l$ increases further the success probability starts to drop.

\begin{figure}[!htb]
\centering
\subcaptionbox{}{\includegraphics[scale=0.5]{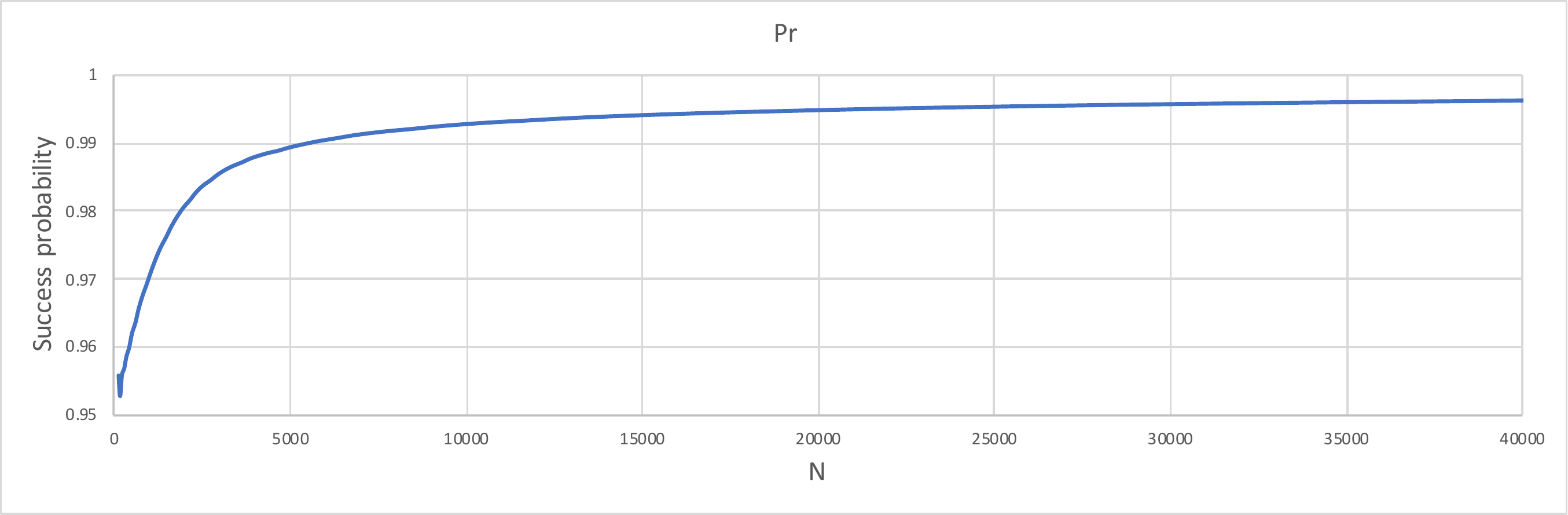}}
\hfill
\subcaptionbox{}{\includegraphics[scale=0.5]{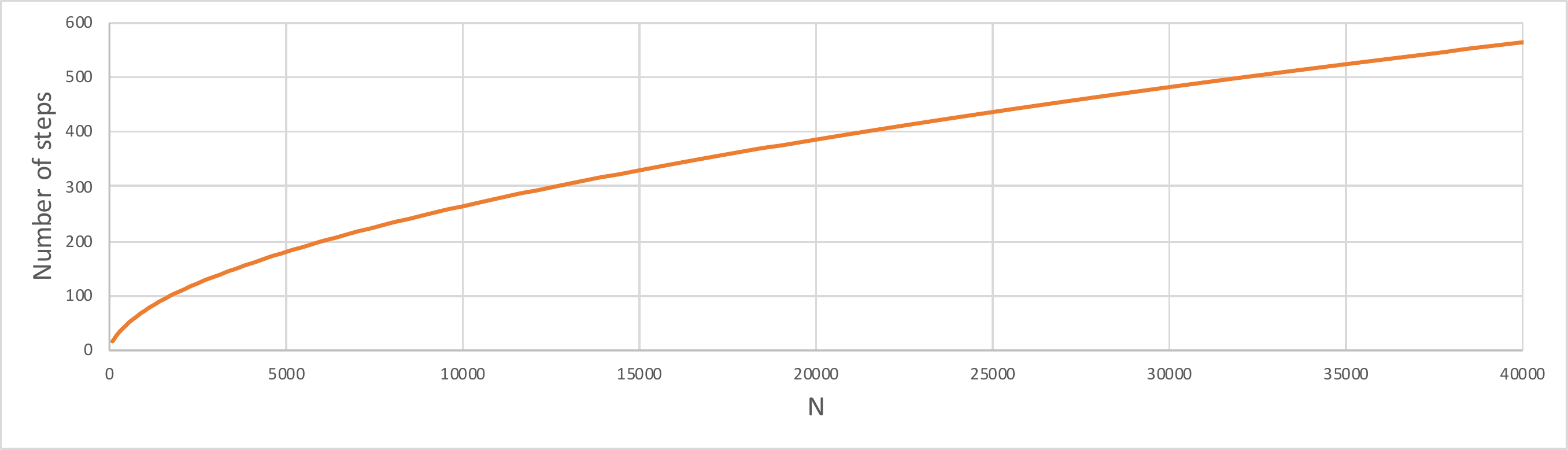}}
\caption{The success probability (a) and the number of steps till the first peak (b) of lackadaisical quantum walk on 2D triangular grid with $l = 6/N$}
\label{fig:tri_diff_n_opt_l}
\end{figure}

Figure \ref{fig:tri_diff_n_opt_l} shows the probability and the number of steps till the first peak for search on triangular 2D grid of $N$ vertices with $l = 6/N$. As one can see as $N$ increases the success probability converges to a constant. The running time (the number of steps till the first peak in this case) fits
  $$
  T = 1.31 \sqrt{N \log{N}} .
  $$

\noindent
Thus, the numerical  simulations suggest that the running time is $O(\sqrt{N \log{N}})$, which is an $O(\sqrt{\log{N}})$ improvement over the loopless algorithm.


\subsection{Lackadaisical quantum walk on honeycomb 2D grid}

\begin{figure}[!htb]
\centering
\includegraphics[scale=0.5]{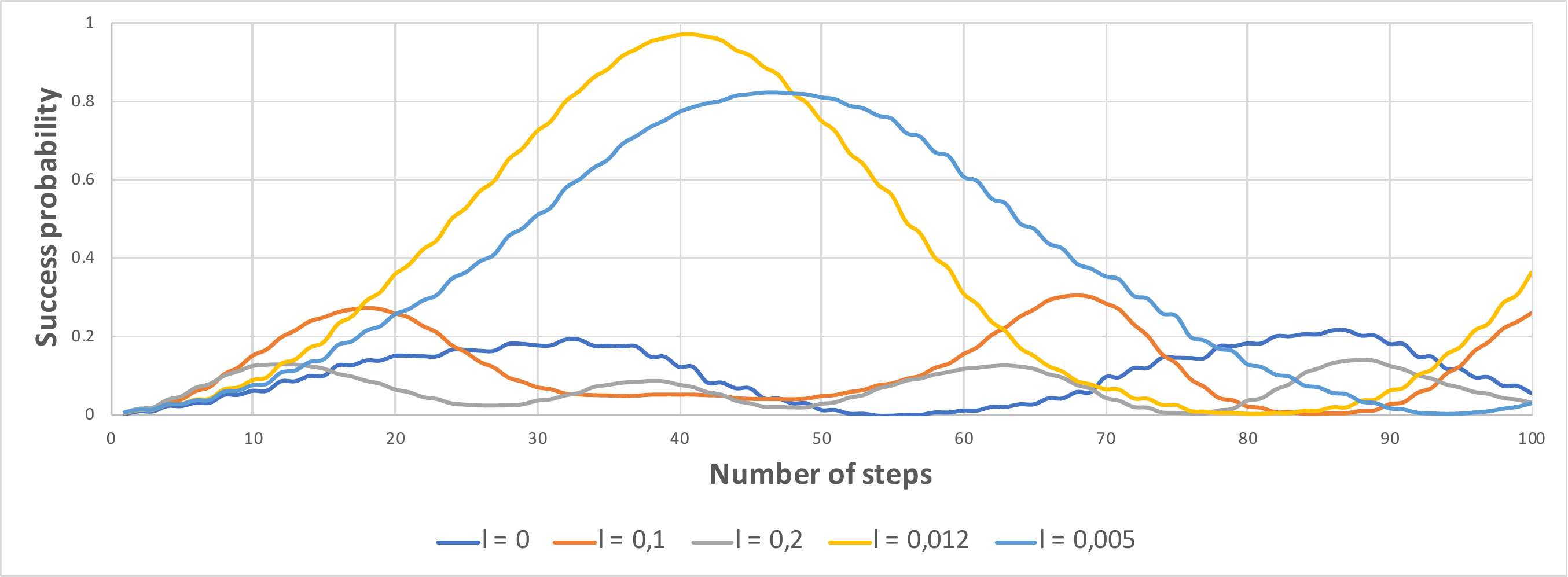}
\caption{Success probability as a function of time for different weights of the self-loop for hexagonal 2D grid of size $16 \times 16$.}
\label{fig:hex_diff_l}
\end{figure}

Figure \ref{fig:hex_diff_l} shows the evolution of the probability of finding a marked vertex for the lackadaisical quantum walk on a hexagonal grid of size $N = 16 \times 16$ for various values of $l$. As one can see different values of $l$ result in different success probabilities and numbers of steps till the first peak.

\begin{figure}[!htb]
\centering
\includegraphics[scale=0.5]{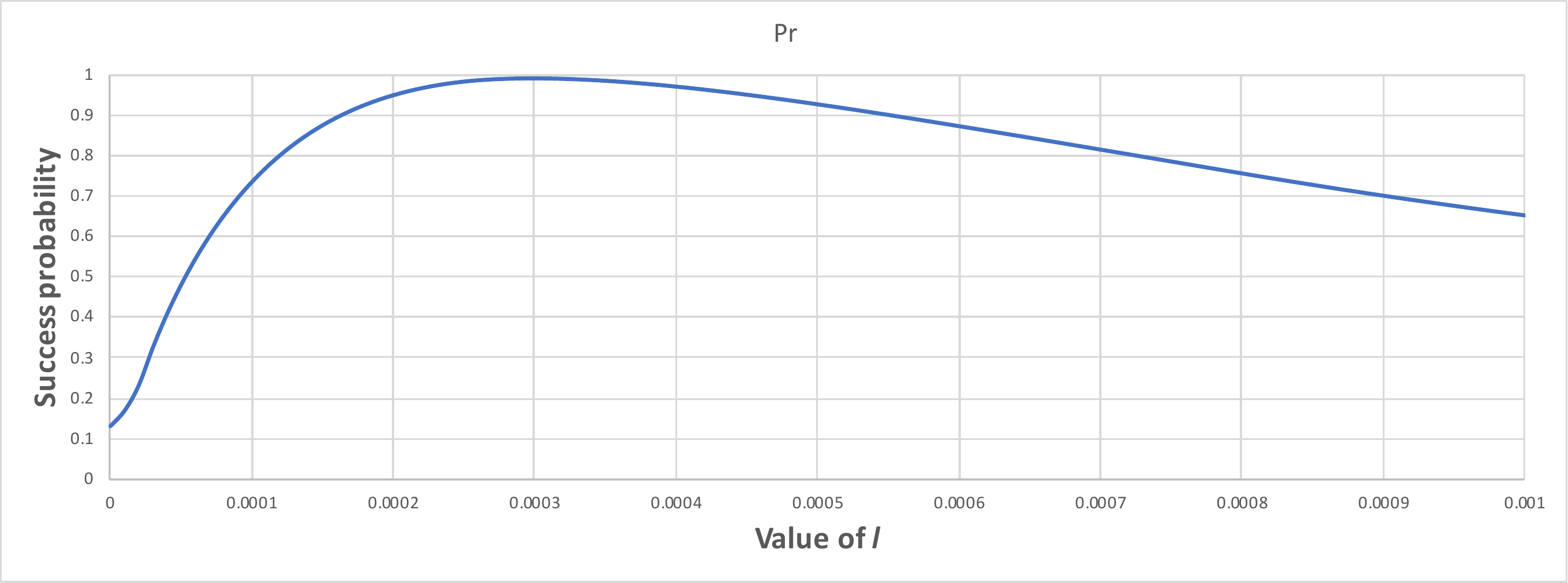}
\caption{Success probability for different weights of the self-loop for hexagonal 2D grid of size $100 \times 100$.}
\label{fig:hex_opt_l}
\end{figure}

Figure \ref{fig:hex_opt_l} shows the success probability for different values of $l$ for search on a hexagonal grid of size $100 \times 100$.

\noindent
When $l = 0$, the lackadaisical walk reproduces the regular (non-lackadaisical) quantum walk, where the success probability reaches a value of $O(1/\log{N})$ at $O(\sqrt{N\log{N}})$. As $l$ increases, the success probability grows, almost reaching $1$ when $l \approx 0.0003$ which is approximately $3/N$. As $l$ increases further the success probability starts to drop.

\begin{figure}[!htb]
\centering
\subcaptionbox{}{\includegraphics[scale=0.5]{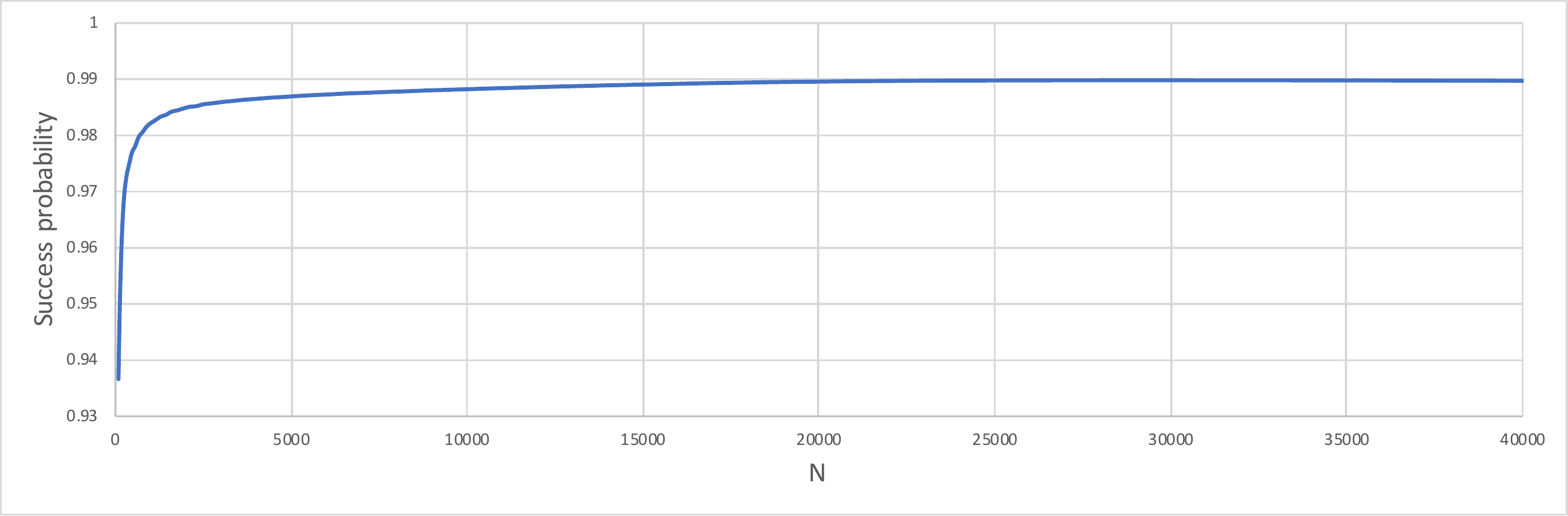}}
\hfill
\subcaptionbox{}{\includegraphics[scale=0.5]{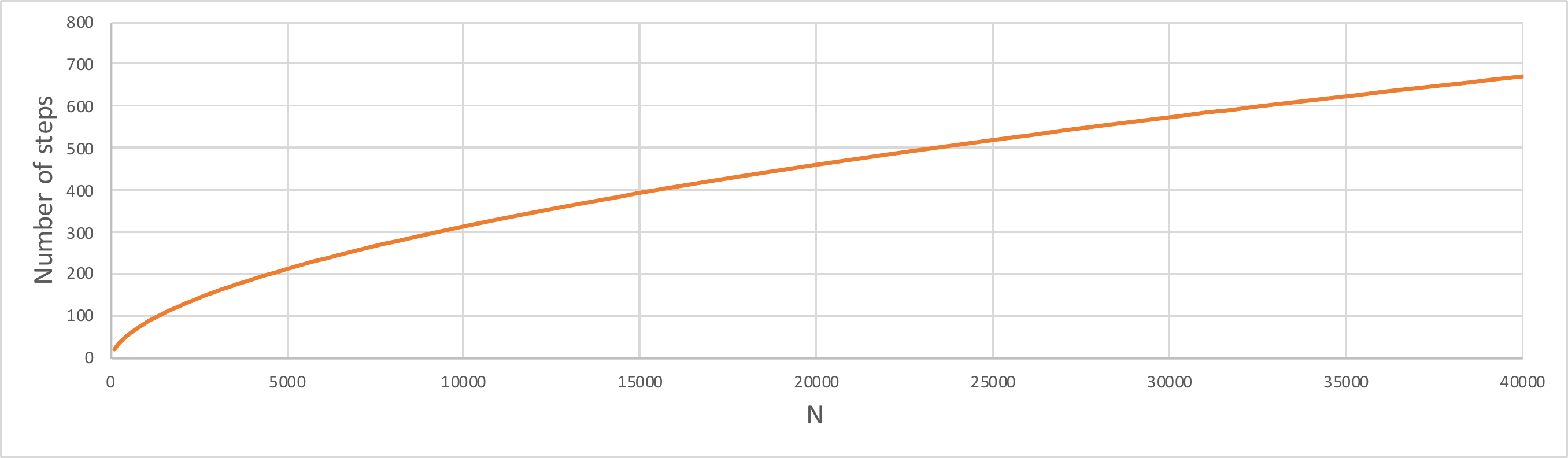}}
\caption{The success probability (a) and the number of steps till the first peak (b) of lackadaisical quantum walk on 2D hexagonal grid with $l = 3/N$}
\label{fig:hex_diff_n_opt_l}
\end{figure}

Figure \ref{fig:hex_diff_n_opt_l} shows the probability and the number of steps till the first peak for search on triangular 2D grid of $N$ vertices with $l = 6/N$. As one can see as $N$ increases the success probability converges to a constant. The running time (the number of steps till the first peak in this case) fits
  $$
  T = 1.56 \sqrt{N \log{N}} .
  $$

\noindent
Thus, the numerical  simulations suggest that the running time is $O(\sqrt{N \log{N}})$, which is an $O(\sqrt{\log{N}})$ improvement over the loopless algorithm.


\section{Conclusions}\label{sec:conclusions}

In this paper, we have studied the search for a single marked vertex by a lackadaisical quantum walk on triangular and honeycomb two-dimensional grids.
We have shown that adding a self-loop, similarly to rectangular grid case, results in $O(\sqrt{N \log{N}})$ improvement over loopless algorithm.
The weight of the self-loop is $6/N$ for triangular grid, $4/N$ for rectangular grid (shown by Wong in~\cite{Wong:2018}) and $3/N$ for honeycomb grid. In all cases the constant in numerator is equal to a degree of a vertex of the grid. 
This observation gives a natural generalisation of search by lackadaisical quantum walk to general graphs, which is the subject for further research.


\comment {
\subparagraph*{Acknowledgements.}
}


\bibliography{Paper}


\end{document}